\documentclass[12pt,preprint]{aastex}
\usepackage[onecolumn,numberedappendix]{emulateapj5}

\usepackage{epsfig}
\usepackage{rotating}
\tightenlines

\newcommand{\epsth}     {\epsilon_{\rm th}}

\newcommand     \bfv  {{\bf v}}
\newcommand     \bbB  {\overline{\bf B}}

\newcommand     \ts  {\times}
\newcommand     \bfb  {\bf b}
\newcommand     \lb{\langle}
\newcommand     \rb{\rangle}
\newcommand     \curl{\nabla {\ts}}

\newcommand\bB{\overline { B}}

\newcommand \lsim{\mathrel{\rlap{\lower4pt\hbox{\hskip1pt$\sim$}}
    \raise1pt\hbox{$<$}}}
\newcommand \gsim{\mathrel{\rlap{\lower4pt\hbox{\hskip1pt$\sim$}}
    \raise1pt\hbox{$>$}}}

\newcommand{\smyr}{{ M_\odot\ \rm yr^{-1}}}
\newcommand{\sm}{{ M_\odot}}

\newcommand{\beq}{\begin{equation}}
\newcommand{\eeq}{\end{equation}}
\newcommand{\beqa}{\begin{eqnarray}}
\newcommand{\eeqa}{\end{eqnarray}}

\newcommand{\mds}{\dot m_*}

\newcommand{\ecore}{\epsilon_{\rm core}}
\newcommand{\esd}{\epsilon_{*d}}

\newlength{\figwidth}
\countdef\decade=200
\advance\decade by \year
\countdef\hours=201
\advance\hours by \time
\divide\hours by 60
\countdef\mins=202
\advance\mins by \hours
\multiply\mins by 60
\multiply\hours by 100
\countdef\miltime=203
\advance\miltime by \hours
\advance\miltime by \time
\advance\miltime by -\mins

\begin{document}

\title{Protostellar Disk Dynamos and Hydromagnetic Outflows\\ in Primordial Star Formation} 


\author{Jonathan C. Tan$^1$ and Eric G. Blackman$^2$
        }
\affil{1. Princeton University Observatory, Peyton Hall, Princeton, NJ
08544, USA.\\jt@astro.princeton.edu}
\affil{2. Dept. of Physics \& Astronomy \& Lab. for Laser Energetics, Univ. of Rochester, Rochester, NY 14627, USA.\\blackman@pas.rochester.edu}

\begin{abstract}
  Are magnetic fields important in primordial star formation? Assuming
  that star formation occurs via an accretion disk that is turbulent,
  initially because of local gravitational instability, 
we calculate the disk
  structure for realistic accretion rates. We predict that local
  gravitational viscosity is able to drive accretion, without the disk
  fragmenting.  We then estimate the rate of dynamo amplification of
  seed magnetic field.  Turbulence in a stratified disk can be
  helical, with different signs of the helicity in each hemisphere.
  This provides a key ingredient for production of global scale
  magnetic fields whose sign of flux is sustained over many orbit
  times. 
  The resulting fields can drive collimated protostellar
  outflows that reduce the star formation efficiency from the initial
  gas cloud, especially once the protostar has contracted to the main
  sequence, at $\sim 100\sm$. We estimate that the outflows are
  powerful enough to eject some material from the host dark matter
  halo and to initiate relatively strong magnetization of the local
  intergalactic medium.  Close to the protostar, the outflow acts to
  shield the disk and equatorial regions from radiative feedback, such
  as ionizing photons, and this may enable accretion up to relatively
  large stellar masses. We conclude that magnetic fields cannot be
  ignored from models of primordial star formation.
\end{abstract}


\section{Introduction\label{S:intro}}

Primordial star formation has attracted much attention recently, due
to advances in numerical models of cosmic structure formation that can
resolve down to solar system scales (Abel, Bryan, \& Norman 2002,
hereafter ABN), constraints on the reionization epoch from the
{\it Wilkinson Microwave Anisotropy Probe} observations of the microwave
background (Kogut et al. 2003), and the discovery of very metal poor
Galactic halo stars (Aoki et al.  2002; Christlieb et al.
2002), whose elemental abundances may be the fingerprints of the
earliest supernovae.

The typical mass of the first stars is an important, but unknown,
quantity. It may be set by feedback processes local to the forming
star. To investigate this possibility it is necessary to model the
star formation process by including as much physics as possible,
whilst making allowance for a realistic geometry of the accreting gas.
Magnetic fields play crucial roles in present-day star formation, but
have usually been assumed to be unimportant in the primordial case. It
is the goal of this paper, which builds on the detailed, yet
non-magnetic, star formation model of Tan \& McKee (2003a), to
critically examine this assumption.
     
At typical locations in the standard cosmological model, baryonic
over-densities collapse to form the first stars at redshifts $\sim 20$
(ABN, Bromm, Coppi, \& Larson 2002). These first ``complex'' baryonic
systems, consisting of a protostar and accretion disk, live for many
dynamical times so that amplification of seed magnetic field by dynamo
action may be possible.  If the field strength becomes dynamically
significant, then an outflow can be generated.

The ubiquity and energy of molecular outflows from present-day
protostars (Zuckerman, Kuiper, \& Kuiper 1976; Kwan \& Scoville 1976;
Edwards \& Snell 1982; Bally \& Lada 1983) was a surprise and forced a
radical change in theoretical models of star formation.  Outflows are
now thought to play a key role redistributing angular momentum in the
accretion disk (Blandford \& Payne 1982), setting the star formation
efficiency (Levreault 1984; Myers et al.  1988; Matzner \& McKee
2000), and depositing energy into the star-forming interstellar medium
(Norman \& Silk 1980). It is often speculated that magnetic fields and
outflows may be unimportant for primordial star formation because the
strength of seed fields is presumably very weak. Thus they have
generally been ignored in studies of the collapsing gas cloud (e.g.
Haiman, Thoul, \& Loeb 1996; Tegmark et al. 1997; Nakamura \& Umemura
1999; ABN; Bromm et al. 2002) and in models of the formation of the
protostellar core (Stahler et al. 1986; Omukai \& Nishi 1998;
Ripamonti et al. 2002; Tan \& McKee 2003a). However, as in the local
Universe, the presence or absence of dynamically-important magnetic
fields and outflows from primordial protostellar disks has important
implications for the basic star formation process.

The amplification of seed magnetic field by dynamo action in
Population~III protostars and its ejection in outflows may also bear
on the long-standing question of the origin of galactic and
intergalactic field strengths (e.g. Kulsrud et al. 1997, and
references therein), particularly if this mode of star formation was
widely and evenly spread throughout the Universe. Such a model would
be analagous to those which propose that fields are ejected from
active galactic nuclei (AGN) (Daly \& Loeb 1990; Colgate \& Li 1999;
Kronberg et al. 2001), though the AGN models would produce more
ordered flux on large scales. If field amplification in
Population~III star formation is very efficient, then much of the
intergalactic magnetic field may have been in place before or at the
same time as the era of reionization, which has implications for
models of field generation in cosmic ionization fronts (Subramanian,
Narasimha, \& Chitre 1994; Gnedin, Ferrara, \& Zweibel 2000).

Starting from conditions representative of the end state of the
numerical simulation of ABN, Tan \& McKee (2003a, hereafter TM)
predicted the mass infall rate, accretion disk structure, and
protostellar evolution associated with primordial star formation.
Using these results, in this paper we consider a wider range of
possible accretion disk models, calculate their structure over the
evolution of the protostar and assess their gravitational stability
with respect to fragmentation (\S\ref{S:disk}). We discuss the ability
of these disks to generate large scale magnetic fields via a dynamo
(\S\ref{S:dynamo}), which is likely to be a key requirement (Lubow et
al. 1994; Blackman 2003; Blackman \& Tan 2003) for magnetically
mediated outflows from thin disks (e.g. Blandford \& Payne 1982;
Lovelace et al. 1987; Pelletier \& Pudritz 1992; K\"onigl \& Pudritz
2000). We show that strong, dynamically important fields can be
produced even from the weak seed fields expected of primordial
protostellar disks. We address the implications of feedback from
hydromagnetic winds on the star formation process in
\S\ref{S:implications}, and conclude in \S\ref{S:conclusions}.

\section{Primordial Protostellar Accretion Disks}\label{S:disk}

The dark matter halos and gas clouds relevant to primordial star
formation form from hierarchical mergers of smaller structures,
including the accretion of matter along filaments (ABN).  Thus in
general we expect that the initial conditions for star formation are
gas cores with non-zero angular momentum. The collapse should proceed
to a disk, which then channels mass to the protostar via viscous
torques.

We use TM's model of the rotating, freely falling accretion envelope
that develops inside the sonic point, $r_{\rm sp}$, of the collapsing
gas. The rotation is parameterized by
\beq
f_{\rm Kep} \equiv \frac{v_{\rm rot}(r_{\rm sp})}{v_{\rm Kep}(r_{\rm sp})} = \frac{v_{\rm rot}(r_{\rm sp})}{(GM_{\rm sp}/r_{\rm sp})^{1/2}}, 
\eeq
which is the ratio of the mass-weighted rotational velocities, $v_{\rm
  rot}$, relative to Keplerian, $v_{\rm Kep}$, of infalling matter at
a radial distance from the protostar corresponding to the
mean location of the sonic point. Inside this region, which
contains a total mass $M_{\rm sp}$, we assume conservation of angular momentum
(Ulrich 1976) so that matter falls freely to the midplane and forms a
disk of size
\beq
\label{eq:rd}
r_d = f_{\rm Kep}^2 r_{\rm sp} \rightarrow 3.44 \left(\frac{f_{\rm
      Kep}}{0.5}\right)^2 \left(\frac{1+f_d}{\esd}\right)^{9/7}\left(\frac{m_*}{\sm}\right)^{9/7} K'^{-10/7}\:{\rm AU}.  
\eeq
For the numerical evaluation we have used the mass-radius relation
expected at the sonic point of the collapsing cloud (TM), which
involves the entropy parameter, $K'$, of the polytropic equation of
state of the cloud, relative to the fiducial case with $T=300\:{\rm
  K}$ at a hydrogen density of $10^4\:{\rm cm^{-3}}$ --- larger values
of $K'$ correspond to denser gas cores. We have also assumed an
efficiency, $\esd$, of collapse from the core to the disk, which is
unity when feedback processes are unimportant, and related $M_{\rm
  sp}$ to the stellar mass, $m_*$, at the time when $M_{\rm sp}$ has
collapsed to the disk: $m_*(1+f_d)=M_{\rm sp}$, where $f_d$ is the
disk mass relative to the star's and with the assumption that there is
little diversion of the accretion flow from $r_{\rm sp}$ to $r_d$.
From the results of the simulations of ABN we have taken a fiducial
value of $f_{\rm Kep}=0.5$, but there is likely to be a significant
dispersion depending on the formation history of the halo and cloud.
The fiducial value of $f_d$ adopted by TM is $1/3$, because the disks
are expected to be relatively massive since their angular momentum
transport must initially be driven by gravitational instabilities.

\begin{figure}[h]
\plotone{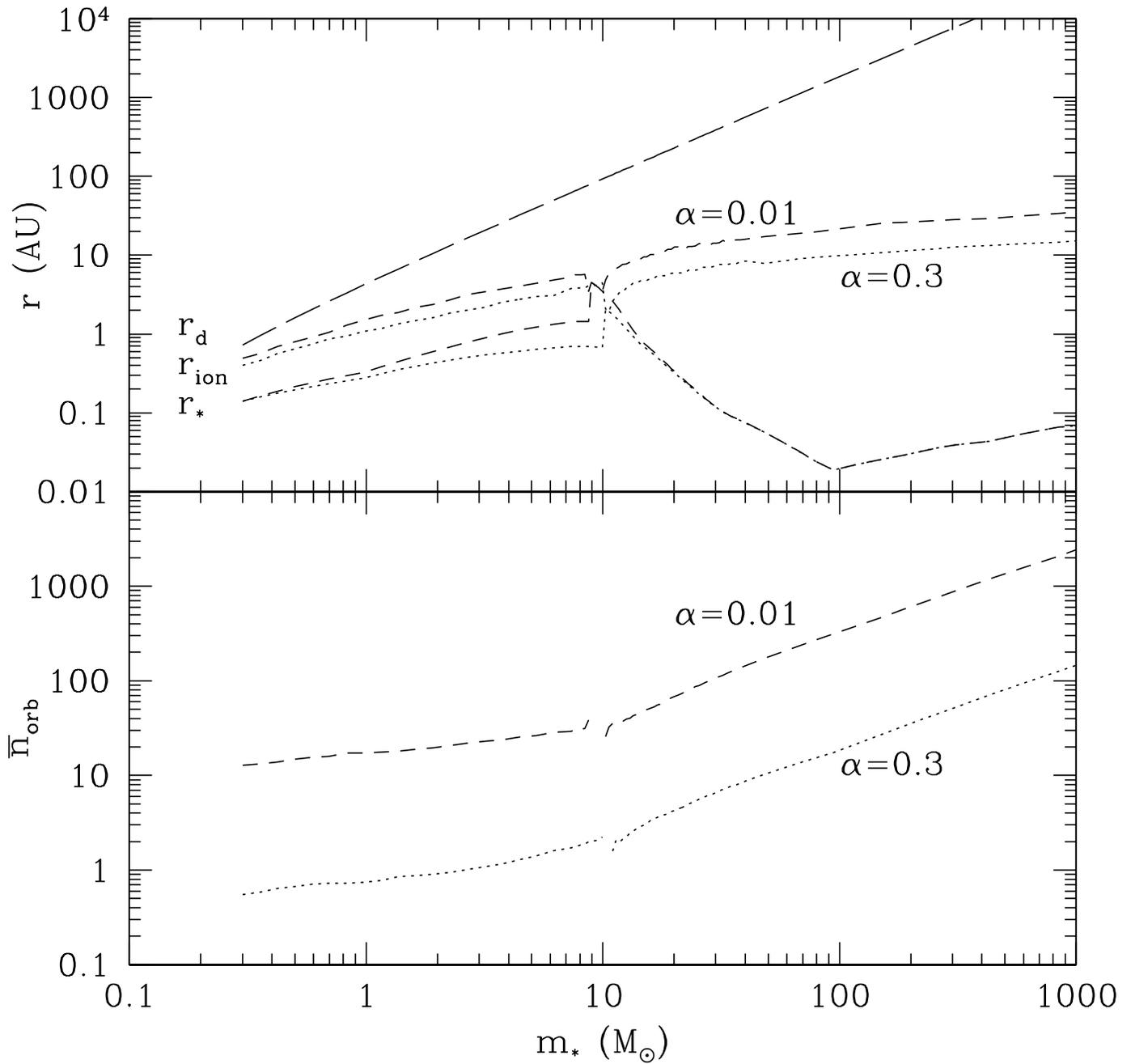}
\caption{
\label{fig:norb}
{\it (a) Upper panel:} Evolution of the radius of the protostar,
$r_*$, and the radial extent of the ionized ($f_i(H)>0.5$) region of
the accretion disk, $r_{\rm ion}$ for models with viscosity parameter
$\alpha_{\rm ss}=0.01,0.3$ (dashed, dotted lines, respectively).
The formation is from the collapse of a core with $K'=1$ and $f_{\rm
  Kep}=0.5$ (see TM for more details).  The full extent of the accretion
disk ($r_d$) is shown by the long-dashed line. 
{\it (b) Lower panel:} Mean number of orbits of fluid elements in
the ionized region of the disk for $\alpha_{\rm ss}=0.01,0.3$ (dashed, dotted
lines). The gap at $m_*\sim 10\sm$ is when $r_{\rm ion}=r_*$ because of the expansion of
the protostar during this stage of the evolution (TM).
}
\end{figure}

Material falls onto the disk (and directly to the star) at all radii
$r<r_d$ (Figure 2 of TM; Cassen \& Moosman 1981). The characteristic
disk scale is much greater than the stellar radius ($r_*\lesssim
100R_\odot=0.5\:{\rm AU}$, TM) so most matter accretes to the star via
the disk. The evolution of $r_*$ as the stellar mass increases is
shown in Figure \ref{fig:norb}a, together with the linear growth of
$r_d$.  Although infall onto the disk provides some effective
viscosity, this is not important in the very inner regions and so we
assume a standard ``$\alpha_{\rm ss}$'' model for the viscosity
(Shakura \& Sunyaev 1973).  We discuss the choices for the value of 
$\alpha_{\rm ss}$ below.

The overall rate of collapse and thus accretion rate of the star is
given approximately by $\mds\sim m_*/t_{\rm ff}$, where $t_{\rm ff}$
is the local free-fall timescale evaluated at the density of the
region that contains mass $m_*$ at the moment it undergoes dynamical
collapse. The accretion rate is thus set by the structure of the
pre-stellar gas core, which in turn is set by the cooling properties
of trace amounts of molecular hydrogen in the primordial gas. A more
careful evaluation of the accretion rate to the star (TM) yields
\beq \mds = 0.026
(1+f_d)^{-10/7}
\esd^{10/7}K'^{15/7}\left(\frac{m_*}{M_\odot}\right)^{-3/7}~~\smyr.
\label{eq:mds}
\eeq
These high
accretion rates make it likely that the disk will build itself up to a
mass significant compared to the stellar mass (we have set $f_d=1/3$) and
will become susceptible to gravitational instabilities.  Two dimensional simulations
of clumpy, self-gravitating disks show self-regulation with
$\alpha_{\rm ss}\simeq (\Omega t_{\rm th})^{-1}$ up to a maximum value
$\alpha_{\rm ss}\simeq 0.3$ (Gammie 2001), where $\Omega$ is the
orbital angular velocity, $t_{\rm th}\equiv \Sigma kT_{\rm c,d}/(\sigma T_{\rm
  eff,d}^4)$ is the thermal timescale, $\Sigma$ is the surface density,
$T_{\rm c,d}$ is the disk's central (midplane) 
temperature, and $T_{\rm eff,d}$ the effective photospheric
temperature at the disk's surface. Fragmentation occurs when $\Omega
t_{\rm th}\lesssim 3$: this condition has the best chance of being
satisfied in the outermost parts of the disk that are still optically
thick (see below). In addition to local gravitational viscosity,
global ($m=1$ mode) instabilities (Adams et al. 1989; Shu et al.
1990) may be efficient at driving inflow if the disk becomes too
massive.

In addition to gravitational instabilities, the magneto-rotational
instability (MRI) may become important, yielding viscous stresses that
correspond to somewhat smaller values of $\alpha_{\rm ss}$. This
process could be particularly important in the inner accretion disk,
where gravitational instability is suppressed. Values often quoted
for the MRI are $\alpha_{\rm ss}\sim 0.01$ (Balbus \& Hawley 1998),
but larger values have also been measured and 
neither simulations nor physical understanding have yet determined
why it cannot differ by an order of magnitude 
(or whether the constant $\alpha_{\rm ss}$ formalism is really appropriate 
for MRI-driven turbulence). We evaluate the ability of the MRI to operate in
\S\ref{S:dynamo}. If the disk is stable with respect to the MRI, then
$\alpha_{\rm ss}$ may be set by molecular viscosity and thus much
smaller. In this case the accretion rate may effectively go to
zero. However, the disk is still being fed at a rate given by equation
(\ref{eq:mds}), so that the surface mass density inevitably builds up
until self-gravity (and the associated larger values of $\alpha_{\rm
  ss}$) becomes important.

Given these considerations, we calculate the radial structure of the
inner disk at any given point in the protostellar evolution by
assuming it is fed smoothly at a rate given by equation
(\ref{eq:mds}). We also use the standard theory of steady, thin,
viscous accretion disks, with a spatially constant viscosity
parameter, $\alpha_{\rm ss}$, and ignore energy injection from the
star. The viscosity is assumed to be a function of the total pressure,
though it is gas pressure that dominates over radiation pressure in
the earlier stages ($m_*\lesssim 20\sm$), which we think are relevant for
field growth (\S\ref{S:dynamo}). 
We evaluate cases with $\alpha_{\rm ss}=0.01,0.3$.

The inner scale of the disk is set by $r_*$, which is predicted from
the protostellar evolution model of TM.  It is somewhat sensitive to
the choice of $\alpha_{\rm ss}$ for $m_*\lesssim 20\sm$ because this
sets the temperature of gas in the inner disk and this thermal energy
is advected into the star (Fig. \ref{fig:norb}a). We adopt primordial
gas abundances $X=0.76$ and $Y=0.24$ so that in the atomic phase
$\mu=1.22m_{\rm H}$ and $n_{\rm He}=0.079 n_{{\rm H}}$, and use the
opacities of zero metallicity gas of Rogers \& Iglesias (1996) for
$T\gtrsim 6000 \: {\rm K}$ and of Lenzuni, Chernoff, \& Salpeter
(1991) for $T\lesssim 6000\: {\rm K}$ (however, here we are interested
primarily in the higher temperature regime). The method assumes that
the disk's central temperature, $T_{\rm c,d}$ is much greater than the
surface temperature, $T_{\rm eff,d}$, i.e. that the disk is very
optically thick: this condition is not well satisfied during the very
earliest stages, or for the outer disk, since the opacity is set by
$\rm H^-$ and declines rapidly for $T\lesssim 7000\:{\rm K}$.

We calculate the ionization state of H and He from the Saha equation
and include the energy needed for ionization of H and He and the thermal 
energy in the disk's energy equation:
\beq
F= \frac{3 G m_*\mds}{8 \pi r^3}
         \left[1-\left(\frac{r_*}{r}\right)^{1/2}\right]
         + \frac{\mds}{4 \pi r} \frac{d}{dr}\left(\frac 53 \bar\epsth
        + \bar\epsilon_I \right),
\label{eq:diskenergyion}
\eeq where $F$ is the flux emitted from one side of the disk, and
$\bar\epsth$ and $\bar\epsilon_I$ are the thermal energy and
dissociation/ionization energy per unit mass, averaged over the
thickness of the disk. The first term on the right hand side of
eq.~(\ref{eq:diskenergyion}) is the viscous dissipation per unit disk
face area.  Note that usually $d\bar\epsth/dr<0$ and
$d\bar\epsilon_I/dr<0$.  The effect of the thermal energy term is
small for thin disks such as those we consider, so we drop it from our
numerical calculations.  However, the dissociation/ionization term,
which is often neglected, can be an order of magnitude larger than
thermal term and can have a major impact on the disk structure. We
therefore retain this term. We follow Frank, King, \& Raine (1995), using an
approximate solution based on radiative diffusion so that $F\simeq
4\sigma T_{\rm c,d}^4/3\tau$, where $T_{\rm c,d}$ is the midplane
temperature and $\tau$ is the optical depth from the midplane to the
surface.

Dividing the disk into many
discrete radial zones, we solve the full set of disk structure 
equations (e.g.  Frank et al. 1995), starting from
the outer regions where the gas is almost completely neutral.  The
high accretion rates of primordial protostars lead to the ionization
energy being quite important, particularly in the earlier stages of
evolution before the protostar starts contracting towards the main
sequence. For complete ionization of atomic gas, the energy absorption
rate is \beq \left( \frac{dE}{dt}\right)_{\rm
  I,max} =\mds\epsilon_{\rm I,max} = 2.4\times
10^{3}\left(\frac{\mds}{10^{-2}\smyr}\right)\:L_\odot,
\label{eq:EI}
\eeq 
calculated assuming a dissociation energy of 13.6~eV for H ionization, 
and 24.6~eV and 54.4~eV for the first and second stages of He ionization.\footnote{The results of ABN indicate that the first solar mass or so of material
  to be accreted would be almost fully molecular, which for a
  dissociation energy of 4.48~eV for $\rm H_2$ increases the
  coefficient of equation (\ref{eq:EI}) by 11\%. We do not include
  this minor correction factor in our numerical calculations.}


\begin{figure}[h]
\plotone{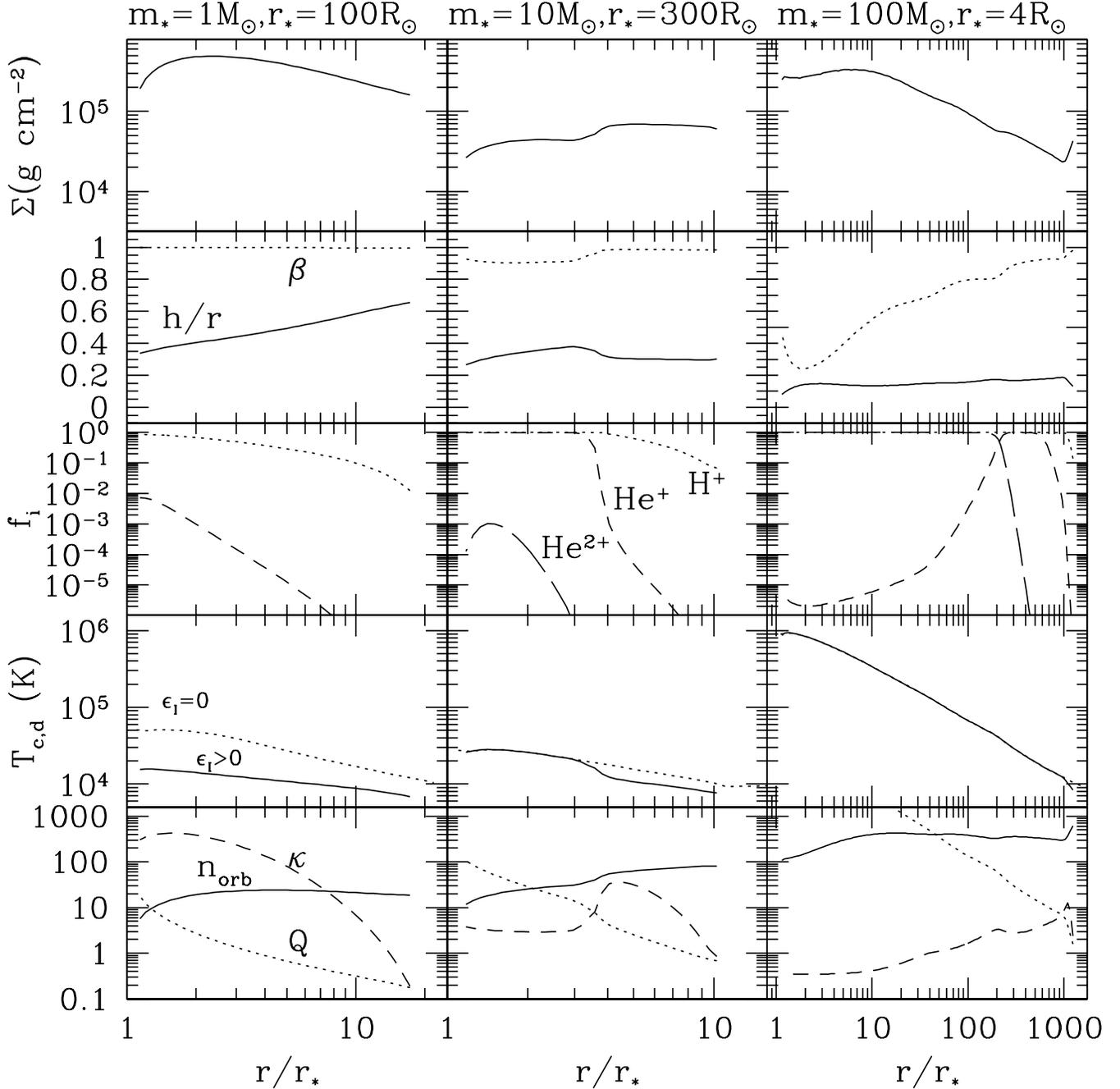}
\caption{
\label{fig:diskexam01}
Protostellar disk structure for models with $\alpha_{\rm ss}=0.01$ and
$m_*=1,10,100\sm$, for which $r_*=100,300,4R_\odot$ and
$\mds=(17,6.4,2.4)\times 10^{-3}\smyr$, respectively. From top to
bottom the panels show (1) surface density, $\Sigma$; (2) ratio of
scaleheight to radius, $h/r$, and ratio of gas pressure to total
pressure, $\beta$; (3) midplane ionization fractions of $\rm
H^+,\:He^+,\:He^{2+}$; (4) disk midplane temperature, $T_{\rm c,d}$ (the
dotted lines show results for when the ionization energy is
neglected); (5) number of orbits, $n_{\rm orb}$, Toomre $Q$ stability
parameter, and Rosseland mean opacity $\kappa$, evaluated at the midplane. 
Note that all
quantities are azimuthal and temporal averages of the disk, which,
being turbulent, exhibits local fluctuations.
}
\end{figure}

\begin{figure}[h]
\plotone{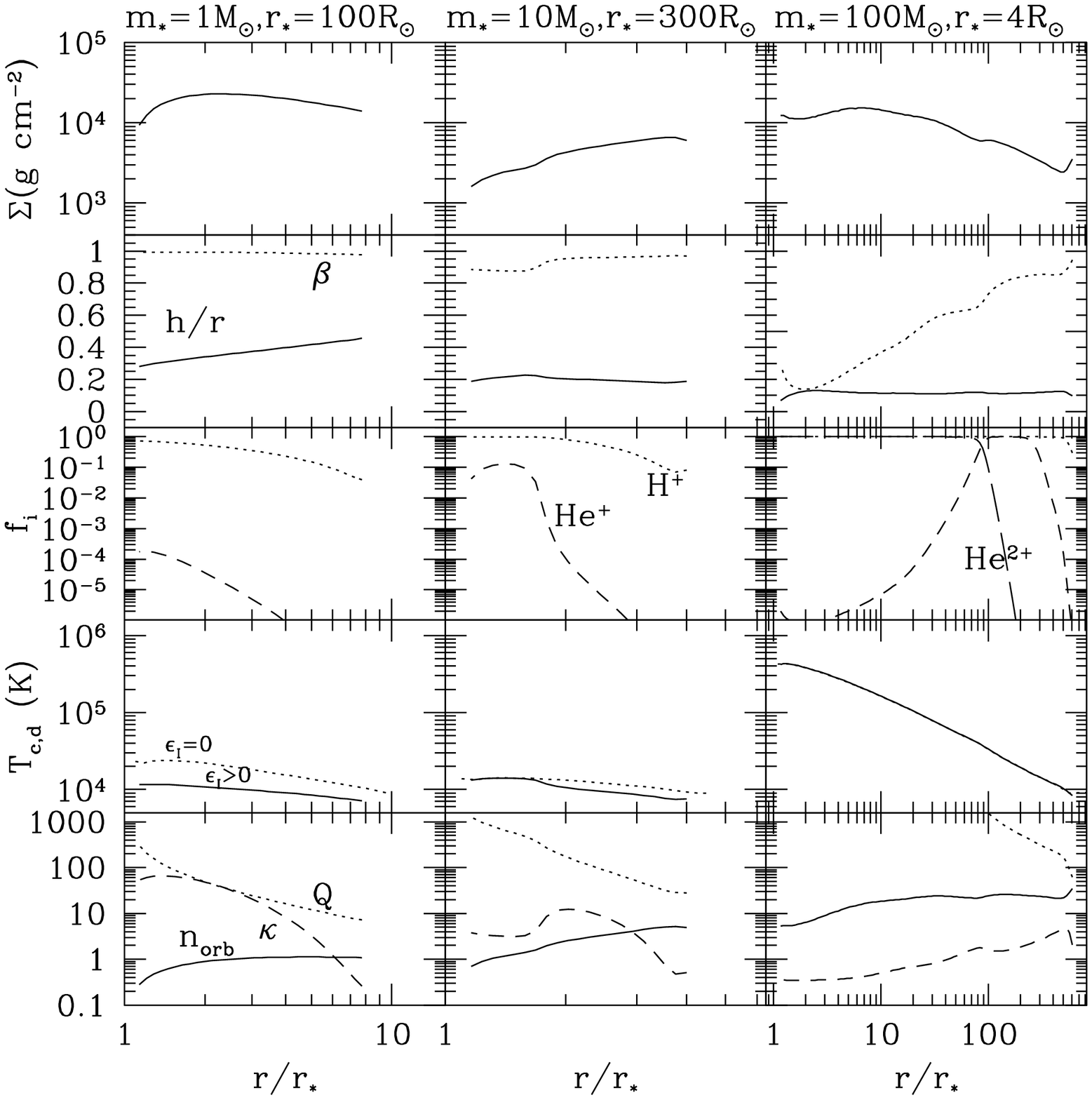}
\caption{
\label{fig:diskexam.3}
As Figure \ref{fig:diskexam01}, but with $\alpha_{\rm ss}=0.3$.
}
\end{figure}

Figure \ref{fig:diskexam01} shows three examples of disk structure for
$\alpha_{\rm ss}=0.01$, with $m_*=1,10,100\sm$. At these masses the
stellar sizes are taken to be $r_*=100,300,4\:R_\odot$ and the
accretion rates (eq. \ref{eq:mds}) are $\mds=(17,6.4,2.4)\times
10^{-3}\smyr$, respectively. Figure \ref{fig:diskexam.3} shows the
equivalent models with $\alpha_{\rm ss}=0.3$. 

The self-gravity in the disks can be gauged by considering the Toomre
stability parameter for a Keplerian disk, $Q\equiv c_s \Omega/(\pi G
\Sigma)$, which is $<1$ in unstable regions. The outer parts of the
disk are most susceptible to gravitational instability. Following a
line of reasoning similar to that presented by Goodman (2003)
concerning self-gravity in quasar accretion disks, consider a disk
with $\alpha_{\rm ss}=0.01$ provided by the MRI: the model will be
physically self-consistent inside the radius at which $Q=1$. At larger
radii, we expect that gravitational viscosity will become
important, allowing larger values of $\alpha_{\rm ss}$ so that $Q$
remains approximately equal to unity.  The results of Gammie (2001)
show that once $\alpha_{\rm ss}\gtrsim 0.3$ then fragmentation occurs.
This corresponds to the location in our $\alpha_{\rm ss}=0.3$ models
where $Q=1$. From Figure \ref{fig:diskexam.3} we can see that $Q>1$
for all the regions of the disks with $T_{\rm c,d}\gtrsim 7000\:{\rm K}$,
and so fragmentation should not occur in these regions. At larger
radii we expect the disks to become optically thin because of the
temperature dependence of $\rm H^-$ opacity (approximately $\propto
T^{-14}$), and have thermal timescales long compared to the orbital
period. This appears to occur before $Q<1$, so that fragmentation
would then not occur anywhere in the disk. 

The infall of material onto the disk provides an effective viscosity
and source of turbulence that is likely to be quite important in the
outer disk, and that this will also help prevent fragmentation. Even
if $Q$ became $<1$, fragmentation is not inevitable: global modes of
gravitational instability, such as spiral waves, could develop
instead. These modes, if present, are relatively efficient at
transporting angular momentum and could then connect the outer infall
region with the inner, viscous accretion disk, where first local
gravitational viscosity, and then perhaps viscosity from the MRI,
bring material all the way to the immediate vicinity of the protostar.
The final accretion to the stellar surface may be mediated in part by
angular momentum loss in protostellar outflows, the properties of
which are discussed in \S\ref{S:implications}.

We conclude that fragmentation of typical primordial protostellar
accretion disks does not occur, if the results of Gammie's (2001)
simulations are applicable. However, further numerical studies are
desirable to confirm this prediction. We note that in the most
advanced simulations to date (ABN) of the global primordial star
formation process, which probe conditions in the initial collapse down
to scales of order astronomical units (just before formation of a
hydrostatic protostellar core and an optically thick accretion disk),
there is also no indication of fragmentation. Primordial star
formation appears to favor single (or perhaps binary) stars rather
that star clusters. The reservoir of gas available to the forming star
is several hundreds to thousands of solar masses, but the typical mass
of the stars is likely to be set by feedback from the protostar acting
on its own accretion. Following TM, here we shall consider the
uninterrupted growth of the star at rate given by eq. (\ref{eq:mds}),
accreting from a stable accretion disk, and use this as the framework
for assessing magnetic field generation and subsequent feedback effects.


\section{Dynamos and Protostellar Disks}\label{S:dynamo}

Stahler et al. (1986) and Omukai \& Palla (2001, 2003) considered the
structure of primordial protostars in the limit of spherical
accretion. They found that the protostar is radiatively stable
(apart from a very thin outer layer, $\sim 10^{-7}$ of the stellar
mass; Omukai 2003, private comm.) until $m_*\simeq 100\sm$, when
nuclear H burning creates a convective core.  Thus, except at very
high masses, little dynamo amplification in the protostar is expected.
We therefore study helical disk dynamos, focusing on the production of
large scale ordered fields because these are the most susceptible to
escape into the corona (Blackman 2003; Blackman \& Tan 2003). There they can drive outflows
that produce a feedback on the star formation process. A non-helical
dynamo may also be present, but this produces disordered random fields
that do not drive a strong outflow. For the helical dynamo to operate,
turbulence in a $stratified$ rotator is the crucial requirement.

\subsection{Basic Theory}

We discuss likely sources of turbulence, the ability of MRI-driven
turbulence to produce large scale fields, and then the helical dynamo.

\subsubsection{Sources of Turbulence}

We expect that the initial driver of turbulence is gravitational
instability (Gammie 2001). This may then be followed by MRI-driven
turbulence (Balbus \& Hawley 1998).

For sufficiently ionized and magnetized disks, the combination of
radially decreasing angular velocity in a rotating plasma and seed
magnetic field can in principle drive the MRI (Balbus \& Hawley 1991,
1998). The vertical stratification then provides the turbulence with
the helical property  from which a large scale helical dynamo feeds.
However, the critical wave number below which the MRI grows, depends
on the initial field strength: there is a minimum initial field for
the instability to drive turbulence in the disk. Since primordial
star-forming environments have only very weak seed fields (discussed
later), it is important to evaluate this critical field strength. It
is determined by requiring that the MRI growth time is shorter than
the diffusion time at the scale of the initial critical vertical
growth wavenumber associated with the initial seed field (Balbus \&
Hawley 1991;1998). This wavenumber is $k_c^2=3\Omega/v_{A}$. Since the
growth rate is of order $\Omega$, the condition that a given
dissipation process is slower than the growth rate requires
\beq
\chi k_c^2/\Omega = 3\chi \Omega/v_A^2 < 1,
\eeq
where $\chi$ is a particular microphysical diffusion coefficient to be
considered.  Note that if $\chi$ is an ambipolar diffusion
coefficient, then no constraint on the field emerges since then
$\chi\propto v_A^2$.  However, if $\chi$ is the Spitzer magnetic
diffusivity (Spitzer 1967), then the condition becomes
\beq
\left(\frac{B}{\rm G} \right)^2 > 4.0 \ts 10^{-4} T_3^{-3/2} \left(\frac{m_*}{\sm}\right)^{1/2} \left(\frac{r}{1000R_\odot}\right)^{-3/2} \left(\frac{{\rm ln}\ \Lambda}{10}\right)^{1/2}
\left(\frac{\rho}{10^{-9}{\rm g\:cm^{-3}}}\right),
\eeq
where $\Lambda$ is the usual Coulomb logarithm and $T_3=T/10^3{\rm K}$.

From this estimate we can see that while the MRI process may be able to
operate 
given the expected initial conditions of present-day star formation,
in the early universe the fields are likely to be too weak: we
estimate in \S\ref{S:expected} that the initial fields that thread the
disk have a maximum value of order $10^{-16}\:{\rm G}$. Yet, even if
the MRI does not work initially, gravitational instability in the disk
can in principle provide local turbulence that is amenable to an
$\alpha_{\rm ss}$ treatment (e.g. Gammie 2001; Wada et al.  2002).  As
we argued in \S\ref{S:disk}, lacking other sources of viscosity, the
accretion disk should establish a state in which the accretion is
driven by gravitational instability. The angular momentum transport
could also be global, via modes that are not describable with the
$\alpha_{\rm ss}$ formalism. The issue of which mechanism operates and
at what levels
is not fully resolved (Balbus \& Papaloizou 1999), requiring further
numerical study.
Here we assume that local turbulence does
arise.

In this context and for the interpretation of simulations, it should
be noted that the meaning of $\alpha_{\rm ss}$ is sometimes
misunderstood: a turbulent disk incurs large fluctuations so that
simulation data must be averaged temporally or spatially for
appropriate comparisons to the $\alpha_{\rm ss}$ formalism to be made.
The formalism for the helical dynamo that we discuss below, only
requires that the stratified helical turbulence yields a viscosity
that can be modeled by an $\alpha_{\rm ss}$ parameter in the mean
statistical sense.

In a turbulent disk, the small scale field grows faster than the
helical large scale field by a factor of order $\alpha_{\rm
  ss}^{-1/2}$ (Blackman 1998, 2003). Thus, well before the large scale
field growth saturates, the small scale field will grow to the
critical value required for the MRI to operate.  For this reason, even
if the initial turbulence is driven by gravitational instability, we
expect the later phase of field amplification to be mediated by the
MRI and its associated $\alpha_{\rm ss}$.


\subsubsection{Comment on MRI and Large Scale Fields}

Field amplification by the MRI in a non-stratified disk is sometimes
said to produce large-scale fields (Balbus \& Hawley 1998), but the
meaning is subtle and easily misunderstood.  In the simplest MRI
simulations that use a non-stratified periodic box, the saturated
magnetic energy spectrum emerges as Kolmogorov with a $k^{-5/3}$
shape.  No global flux can grow in a periodic box. The toroidal field
has the largest coherent scale, but this is only of order the disk
height.  Most importantly, although the energy of this field is
sustained, the sign of the field fluctuates rapidly, on a timescale of
order the turbulent correlation time (about equal to a Keplerian
rotation time for MRI unstable disks).  
The reason is that scales of order the box thickness
are themselves unstable to the MRI, and thus are the largest turbulent
scale. The largest scale fields produced by the non-stratified MRI
have the same spatial and temporal properties of the largest scale of
the turbulence (Hawley et al. 1995).  Such turbulent scale fields can
be shredded before buoyantly rising to the coronae (Blackman 2003;
Blackman \& Tan 2003), and are therefore not ideally suited for
driving outflows.

However when stratification is present, the MRI can produce helical
turbulence.  ``Helical'' describes the property that rising turbulent
eddies in the northern (southern) hemisphere expand and twist
clockwise (counterclockwise) to conserve angular momentum, while
falling turbulent eddies twist counterclockwise (clockwise).  Both
rising and falling plasma maintain the same sign of pseudoscalars such
as $\lb\bfv\cdot\curl\bfv\rb$ and $\lb\bfb\cdot\curl\bfb\rb$, (where
$\bfv$ and $\bfb$ are the turbulent velocity and turbulent magnetic
field in Alfv\'en units, respectively), with each having opposite signs
in opposite hemispheres.  Such finite pseudoscalars enable amplification of global scale
fields by helical dynamo action when the rising plasma is
threaded by a magnetic field (e.g. Moffatt 1978; Parker 1979;
Brandenburg 2001; Blackman \& Field 2002).
In a real non-periodic system the helical dynamo can
maintain a global large scale flux over many rotation periods, as seen
in non-periodic box simulations of disks (Brandenburg et al. 1995;
Brandenburg \& Donner 1997). Note that 
Stone et al. (1996) claimed that no helicity was observed in their
stratified periodic box disk simulations, when averaged over
hemispheres (Stone 2003, private comm.). Further studies are required
to reconcile these results.

\subsubsection{The Helical Dynamo}

Recently, Blackman \& Field (2002) developed a dynamical nonlinear
helical dynamo theory that accounts for the nonlinear backreaction of
the magnetic field on large scale field growth and agrees with recent
helical dynamo simulations (Brandenburg 2001).  The driver of
exponential large scale field growth, $\alpha_d$, turns out to be
proportional to the residual helicity $\lb\bfv\cdot\curl \bfv\rb-
\lb\bfb\cdot\curl\bfb\rb$ 
(Pouquet, Frisch, \& Leorat 1976), 
rather than just the kinetic helicity
$\lb\bfv\cdot\curl\bfv\rb$ used in textbooks for similar dynamos.
Large scale field growth occurs in two phases: an early, fast
kinematic phase, and a late, slow resistive phase.  Generation of the
large scale field is associated with large scale magnetic helicity,
whilst helicity conservation means that the small scale helicity grows
with the opposite sign.  As this happens, $\lb\bfb\cdot\curl\bfb\rb$
grows and offsets the $\lb\bfv\cdot\curl\bfv\rb$, shutting down the
dynamo.  These principles seem to survive in dynamos with shear
(Blackman \& Brandenburg 2002).  Although the above results are for
simplified dynamos, we shall use the saturation principles in what
follows, by incorporating them into an $\alpha-\Omega$ accretion disk.
Other approaches to mean field accretion disk dynamos have been
developed by, for example, Pudritz (1981), R\"udiger, Elstner, \& Stepinski (1995),
Stepinski (1995), Reyes-Ruiz \& Stepinski (1997; 1999) and Campbell \&
Caunt (1999).

To estimate the large scale field, $\bbB$, produced by a helical
dynamo, we need the growth rate. We use a mean field theory for large
scale field growth in a turbulent disk. This approach presumes that
the mean quantities vary on time scales long compared to that of the
fluctuating turbulent quantities.  Therefore, the derived mean field
growth rate, $\gamma$, should be slower than the eddy turnover rate,
i.e. $\gamma < \Omega$, where the latter is also the growth rate of
the random field (e.g. Balbus \& Hawley 1998). We shall find that this
condition is indeed satisfied.

We assume  ${\bbB} = {\bbB}_0 {\rm Exp}[\gamma t +ik_z z]$,
where $k_z$ is the vertical wavenumber.
The basic equations for the radial and toroidal field in   
an 
``$\alpha_d\Omega$'' dynamo for a Kelperian disk 
become
(e.g. Ruzmaikin et al. 1988)
\beq
\gamma{\bB}_r=-ik_z(\alpha_d \bB_\phi) - \beta k^2_z\bB_r
\label{6}
\eeq
and
\beq
\gamma{\bB}_\phi=-(3/2)\Omega \bB_r - \beta k^2_z\bB_\phi,
\label{7}
\eeq
where we ignore spatial dependencies of the 
pseudoscalar $\alpha_d$ and  turbulent transport scalar $\beta$,
and assume averages are taken within a given hemisphere.
We assume $\alpha_d=q\alpha_{d0}$ 
where the time-dependent multiplicative factor $q\le 1$ 
represents the backreaction from $\lb\bfb\cdot\curl\bfb\rb$, 
with $q=1$ in the kinematic regime.

There are two possible dominant contributions to $\alpha_{d0}$: a
Coriolis force effect and a magnetic buoyancy effect.  First consider
the Coriolis force acting on the largest turbulent eddies, of typical
scale $l$.
A rising and expanding (falling and contracting) eddy rotates in the
opposite (same) sense as that of the underlying disk rotation during
its turbulent correlation time, $t_{\rm ed}=l/v$.
Therefore, while conserving angular momentum about the disk axis, such
eddies gain a specific angular momentum about their own symmetry axis
such that $\lb\bfv\cdot\curl\bfv\rb$ is non-vanishing and of opposite
sign in each hemisphere.  Using the equation for $\curl \bfv$ evolution
and the mass continuity equation, and assuming that the density varies
on scale $h$, this leads to the estimate (Ruzmaikin et al. 1988)
\begin{equation}
\alpha_{d0} = t_{\rm ed} \lb{\bf v} \cdot \curl {\bf v}\rb \sim 
t_{\rm ed}^2 \lb {\bfv}\cdot(\curl( \Omega \ts {\bfv}))\rb
\sim t_{\rm ed}^2 \lb {\bfv}\cdot{\bf\Omega} \nabla\cdot\bfv\rb\sim 
{\Omega l^2\over  h}\times {\rm Min}[R_b,1],
\label{alphad}
\end{equation}
where the last relation follows from mass continuity,
and $R_b\equiv {v\over l\Omega}$ is the Rossby number.

For thin disks in the Shakura-Sunyaev formalism, 
\beq
l\sim \alpha_{\rm ss}^{1/2}h/R_b^{1/2}
\label{ldef}
\eeq 
and 
\beq
v\sim \alpha_{\rm ss} ^{1/2}R_b^{1/2}c_s,
\label{vturb}
\eeq
 which follow since $\nu \sim v l\sim v^2/(R_b\Omega)
= \alpha_{\rm ss} c_s h$ and $\Omega h\simeq  c_s$.  
Using (\ref{ldef}) in  (\ref{alphad}), we have
\begin{equation}
\alpha_d = q\alpha_{d0} \sim  q \alpha_{\rm ss} \Omega h.
\label{alphad2}
\end{equation}
It also follows that the maximum fractional kinetic helicity satisfies 
\beq
f_h={\rm Min}[1,
\alpha_{d0}/v \sim \alpha_{\rm ss}^{1/2}/R_b^{1/2}].
\label{fh}
\eeq

It is important to note that while equation~(\ref{alphad}) is a
standard estimate for $\alpha_{d0}$, Brandenburg \& Donner (1997)
found that for large scale field generation in local shearing box
simulations, the contribution seemed to have the opposite sign of that
required to produce the measured mean field.  It was later suggested
(Brandenburg 1998) that a magnetic buoyancy term of the form $t_{\rm
  ed} \lb b_x b_y (g_z /p) \rb$ where $b_x,b_y$ are the fluctuating
components of the magnetic field, $g_z$ is the vertical gravity, and
$p$ is the pressure, actually has the correct sign.  Since this term
is $\sim(1/\Omega)(v_A^2/c_s^2) \alpha_{\rm ss} (v_\phi^2/r) (h/r)
\sim \alpha_{\rm ss} c_s \sim \Omega l^2/h$, which has the same order
of magnitude and scalings as equation~(\ref{alphad}), it does not
significantly affect our calculations based on the use of
equation~(\ref{alphad2}), unless the two cancel to high order.  The
cycle period analyzed by Brandenburg \& Donner (1997) suggests some
cancellation.  However, they studied the late time evolution of the
dynamo, after the kinematic regime saturates.

With our assumed form for $\bbB$, we solve equations (\ref{6}) and
(\ref{7}), use equation~(\ref{alphad2}), and set $k_z\sim 1/h$ and
$\beta\sim \nu=\alpha_{\rm ss}c_s h$, to derive the growth rate
\begin{equation}
\gamma=(3\alpha_d k_z \Omega/4)^{1/2}-\beta k_z^2
=\alpha_{\rm ss}^{1/2}c_s({\sqrt 3}q/2 -\alpha_{\rm ss}^{1/2})/h.
\label{gamma}
\end{equation}
Note that with $q=1$, for growth this requires a dynamo number $D
\equiv |(3/2)\Omega\alpha_0h^3/\beta^2| >2 = D_c$, the critical dynamo
number. Critical dynamo numbers for disks are sometimes derived to be
$>8$ (e.g. Ruzmaikin et al. 1988) but this value depends on the
assumed profile of $\alpha(z)$.  We have crudely assumed a nearly
constant profile with $z>0$, and thus our critical dynamo number is
low.


At a given stage in the evolution when the stellar mass is $m_*$, we
estimate $\bB$ as the smaller of (i) the saturation value at the end
of the kinematic regime and (ii) the value reached given the growth
time that the disk dynamo has been able to operate, 
\beq
t_{\rm grow} = t(m_*) - t_{\rm init},
\eeq where $t_{\rm init}$ is the time at which the
protostellar disk becomes susceptible to turbulence. We define $t=0$
at the time when the protostellar core has just started to form, then
from equation~(\ref{eq:mds}) we have 
\beq
t(m_*)=27(1+f_d)^{10/7} \esd^{-10/7} K'^{-15/7} (m_*/M_\odot)^{10/7} \:{\rm yr}.
\eeq
Several factors potentially determine $t_{\rm init}$. Firstly, we require a
critical level of ionization so that the field's growth time due to
the dynamo is substantially shorter than the diffusion time.
Brandenburg et al. (1995) found a critical fractional ionization of
disk material of $\sim 10^{-14}$ for growth of the MRI, corresponding
to a neutral-ion collision frequency $\sim 10$ times the orbital
frequency.  The MRI has a growth time $\alpha_{\rm ss}$ times that of
the large scale helical field and so our critical ionization fraction
is $\sim \alpha_{\rm ss}^{-1} 10^{-14} \sim 10^{-13}$.  This
condition is well-satisfied for the inner regions of primordial
protostellar disks, once the star has a mass that is even just a small
fraction of a solar mass.  Thus $t_{\rm init}$ (and the corresponding
$m_{\rm *,init}$) is set by the time for disk formation in the
collapsing flow, i.e. when the centrifugal radius of the flow becomes
significantly larger than the protostellar radius.  For the fiducial
case of $K'=1$ and $f_{\rm Kep}=0.5$, this occurs very early on, when
the star has mass $m_{\rm *,init}\simeq 0.4\sm$.

One might think that the time available for field growth at a given
radius is the accretion timescale at that radius, rather than the disk
age.  However, since the mean field formalism is statistical,
amplification occurs as long as the system remains in
quasi-statistical equilibrium (i.e. as long as helically turbulent
plasma is steadily supplied to that location).  A related point is
that as long as $\beta > (L^2/r^2) \nu$, where $L\sim h$ is the
vertical variation scale of the large scale field, a poloidal mean
field is likely not frozen into the plasma on an accretion timescale
and would diffuse away if the $\alpha_d$ helicity effect is ignored.
This then motivates the need for the $\alpha_d$ effect and large scale
dynamos in the first place (Lubow et al. 1994; Blackman 2003; Blackman
\& Tan 2003).  For completeness however, we will also provide the
estimated field when the growth time is limited to the local accretion
time.

After the kinematic regime, subsequent field growth becomes
resistively limited as the kinetic and current helicities nearly
balance. The magnitude of the helical component of the mean field
when this occurs is given by (Blackman \& Field 2002; Blackman 2003)
\beq
\bB_H \sim (4\pi \rho f_h)^{1/2} v (l/h)^{1/2} \simeq 
 (4\pi \rho)^{1/2} \alpha_{\rm ss} c_s,
\label{sat}
\eeq where the latter similarity follows from equations (\ref{ldef}),
(\ref{vturb}), and (\ref{fh}).  We take this to be an estimate of the
poloidal field, i.e. we assume $\bB_r\sim \bB_H$. We assume that the
strength of the toroidal field at the surface $\bB_\phi$, is
comparable to its value within the disk.  This toroidal field is
larger than $\bB_r$ because it is linearly amplified above the value
of $\bB_r$ by shear as it rises into the coronae during a buoyancy or
diffusion timescale.  These timescales are $t_{\rm buo}\sim h (4\pi
\rho)^{1/2}/\bB_\phi$ and $t_{\rm dif}=h/\alpha_{\rm ss} c_s$,
respectively.  Linear growth of $\bB_\phi$ above the level predicted
by equation~(\ref{sat}) in a time $t_{\rm buo}$ gives $\bB_\phi=\bB_r
\Omega t_{\rm buo}=\alpha_{\rm ss}^{1/2}c_s (4\pi\rho)^{1/2}$, which
from equations (\ref{sat}) and (\ref{vturb}) implies $\bB_r/\bB_\phi=
\alpha_{\rm ss}^{1/2}$. Similarly, if the growth of $\bB_\phi$ is
limited by turbulent diffusion, then $\bB_\phi=\bB_r \Omega t_{\rm
  dif}$, so that $\bB_r/\bB_\phi \sim \alpha_{\rm ss}$.

To summarize: for an initial seed field $\bbB_0$ 
we expect the final field strengths to be 
\beq
\bB_r \simeq {\rm Min} \left\{ \bB_{0}{\rm Exp} [\gamma_0 t_{\rm grow}],
\ \ \alpha_{\rm ss} c_s (4\pi \rho)^{1/2} \right\},
\label{br}
\eeq
where $\gamma_0\simeq \alpha_{\rm ss}^{1/2} c_s/h=\alpha_{\rm ss}^{1/2}\Omega$ is the kinematic $\gamma$, obtained from 
equation~(\ref{gamma}) with $q\sim 1$, 
and  
\beq
\bB_\phi = \alpha_{\rm ss}^{-1}\bB_r 
\eeq
if $\bB_\phi/(4\pi \rho)^{1/2} < \alpha_{\rm ss}c_s$,
(i.e. when $t_{\rm buo}> t_{\rm dif}$)
or 
\beq
\bB_\phi = \alpha^{-1/2}_{\rm ss} \bB_r
\label{bphi2}
\eeq if $\bB_{\phi}/(4\pi \rho)^{1/2} > \alpha_{\rm ss}c_s$, (i.e.
when $t_{\rm buo}< t_{\rm dif}$). The latter case applies in the
saturated state.

\subsection{Expected Values for Star-Forming Disks\label{S:expected}}

To determine the expected $\bB$ in star-forming disks, we must
estimate the seed field, assess flux freezing for the initial collapse
of the cloud to form the disk, and estimate the growth timescale to
determine if the field has saturated.
The seed field for the primordial star-forming environment can be
estimated from the Biermann battery mechanism for non-barotropic
flows.  
On scales much larger than that of an indvidual star, 
numerical simulations indicate that by $z=18$, seed fields of
order $\bB_0\sim 10^{-26}$ Gauss can be generated (Kulsrud et al.
1997).  We assume such a field is generated inside the pre-stellar
core region in the simulation of ABN, which has a size $\sim 100
\:{\rm pc}$ and a density $\sim 10^{-26}\:{\rm g\:cm^{-3}}$. The
infall speeds within this region are of order km/s, vastly greater
than the drift speeds of the ions through the neutrals (the ionization
fractions are much greater than the equilibrium values for the gas,
which has a minimum temperature of $\sim 200\:{\rm K}$). Thus field is
frozen into the contracting gas: going from parsec to AU scales the
field strength increases by factors of $\sim 10^{10}$, so the seed
field in the disk may then be $\sim 10^{-16}\:{\rm G}$.

It may actually be that much larger seed fields 
could arise in the protostar from the Biermann battery term
if the pressure and density gradients are misaligned inside the protostar
as well. The protostellar Biermann battery (e.g. Kulsrud et al. 1997) value
reached after a rotation time (at which point other 
growth terms become important) can be
estimated to be ${\overline B}_0\simeq (m_H c/e){\nabla P \ts \nabla\rho\over \rho^2 \Omega}
\sim 2 \ts 10^{-11}(T/10^5{\rm K})^{1/2}(r/10^{13}{\rm cm})^{-1} (h/r)$~Gauss,
where we have used $\Omega=c_s/h$ and have 
conservatively taken $\nabla \sim 1/r$.

In the disk, flux freezing and field advection are not guaranteed
since, as discussed above, turbulent diffusion of the mean field may
be relatively efficient and the accretion timescale may be relatively
long. Thus, in addition to a fiducial initial seed field strength of $\bB_0=10^{-16}{\rm G}$, we also consider a conservative value of
$\bB_0 =10^{-26}{\rm G}$. Since the disk-dynamo amplifies the
initial field exponentially, the uncertainties in the strength of the
seed field do not significantly affect the time to reach saturation.

\begin{figure}[h]
\plotone{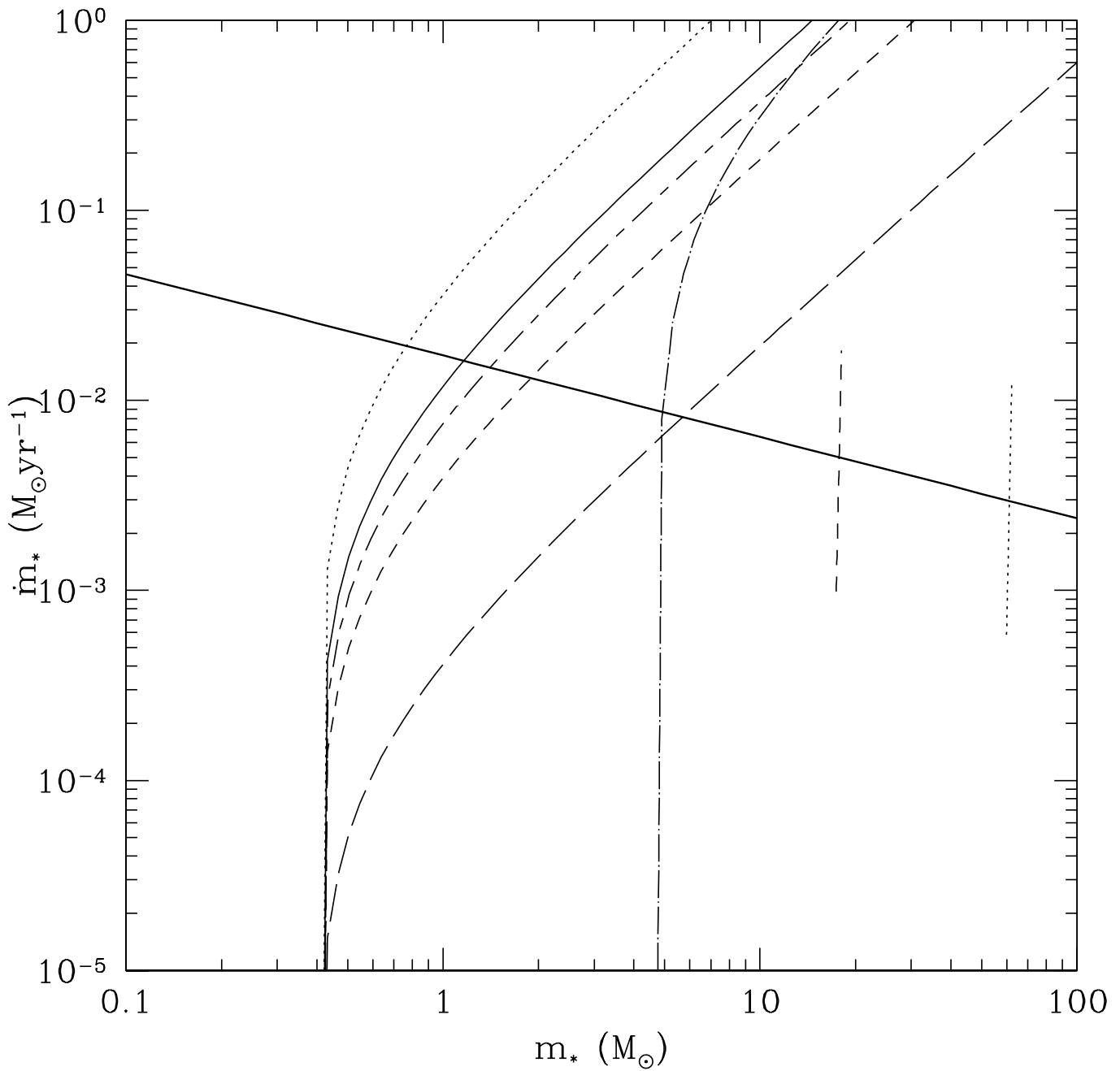}
\caption{
\label{fig:param}
\footnotesize Conditions for saturation of the disk magnetic field.
Protostellar growth starts from low masses and with accretion rates
that decline with time and mass: the fiducial case is shown by the
straight thick solid line. The rising curves represent the condition
that the disk magnetic field has been able to saturate to values given
by equation (\ref{exp2}), when the disk is old enough to have allowed
sufficient amplification of the seed field of $10^{-16}\:{\rm G}$. The
disk dynamo is assumed to start operating at $m_{\rm *,init}=0.4\sm$,
which corresponds to when the centrifugal radius of the accretion flow
is about an order of magnitude greater than the stellar radius in the
fiducial case with $f_{\rm Kep}=0.5$.  The dashed, solid, and dotted
lines show the saturation condition for $\alpha_{\rm
  ss}=0.001,0.01,0.1$, all with $r=10^{13}\:{\rm cm}$ and $m_{\rm
  *,init}=0.4\sm$.  With $\alpha_{\rm ss}=0.01$, the long-dashed line
shows this condition for $r=10^{14}\:{\rm cm}$, the dashed-long-dashed
line shows it for a seed field of $10^{-26}\:{\rm G}$, and the
dot-long-dashed line shows it evaluated with $m_{\rm *,init}=4.7\sm$,
corresponding to an initial core with $f_{\rm Kep}=0.1$. If the dynamo
amplification time is limited by the accretion timescale then
saturation occurs at masses indicated by the short line sections
(dashed for $\alpha_{\rm ss}=0.01$ and dotted for $\alpha_{\rm
  ss}=0.3$), which use the results for $\bar{n}_{\rm orb}$ of
\S\ref{S:disk} and assume a seed field of $10^{-16}\:{\rm G}$. Note
that these results for $\bar{n}_{\rm orb}$ are only valid for $m_*$
and $\mds$ close to the fiducial case shown by the thick solid line.
}
\end{figure}

Using $\gamma_0\sim \alpha_{\rm ss}^{1/2} c_s/h=\alpha_{\rm
  ss}^{1/2}\Omega$ from (\ref{gamma}) with $q=1$, and approximating
$t_{\rm grow} = t(m_*)=0.7 m_*/\mds$, the
first term on the right of equation~(\ref{br}) becomes
\beq
\bB_r \sim 8.9\times10^8 {\rm G}
\left(\frac{\bB_0}{10^{-26}{\rm G}}\right){\rm Exp}\left[\left(\frac{\alpha_{\rm ss}}{0.01}\right)^{1/2}\left(\frac{m_*}{M_\odot}\right)^{3/2}
\left(\frac{r}{10^{13}{\rm cm}}\right)^{-3/2}
\left(\frac{{\dot m}_*}{0.01 M_\odot {\rm yr}^{-1}}\right)^{-1}\right],
\label{exp}
\eeq
while the second term gives
\beq
\bB_r \sim 49 {\rm G}(\alpha_{\rm ss}/0.01)^{1/2}(m_*/M_\odot)^{1/4}
({\dot m}_*/0.01M_\odot {\rm yr})^{1/2}
(r/10^{13}{\rm cm})^{-3/4}(h/10^{12}{\rm cm})^{-1/2},
\label{exp2}
\eeq where mass continuity at $r=2r_*$, $\rho= {\dot m}_*/(4\pi h^2
\alpha_{\rm ss} c_s)$, was used, and we choose parameters commensurate
with the results of \S\ref{S:disk}. Although primordial protostars are
thought to have accretion rates somewhat higher than the
$10^{-2}\smyr$ adopted in equation~(\ref{exp}), it is clear from these
equations that we expect the field to saturate when the protostar has
a relatively small mass. Note that the seed fields for present-day
Galactic protostellar disks would be many orders of magnitude higher
than the primordial case: the initial Galactic interstellar material
possesses a field of $\gtrsim 10^{-6}$ G.  Thus for a wide range of
parameters equation~(\ref{exp2}) is the relevant estimate for the
dynamo-amplified large scale field strengths in present-day, as well
as primordial, protostellar disks. For the primordial case, we
evaluate the saturation point more accurately in the $m_*$ versus
$\mds$ plane, and show the results in Figure~\ref{fig:param}.

In the fiducial case the field in the inner disk saturates by
about $1\sm$, and this does not depend very sensitively on
$\alpha_{\rm ss}$. In lower angular momentum cores the disk emerges
later, at which point the dynamo rapidly causes the field to saturate.
Although we expect that mean field dynamo can grow field at a given
location for longer than the local accretion time, we also evaluate
the cases where $t_{\rm grow}=t_{\rm acc} \sim 2\pi n_{\rm orb}/\Omega
\simeq r^2 /(\alpha_{\rm ss} c_s h)$. The resulting field strength
must always be smaller than that from equation~(\ref{exp}) for disks
for which $\alpha_{\rm ss}$ models are self-consistent.  The
saturation point is delayed until masses $m_*\sim 10-20\sm$ in the
fiducial case.




\section{Implications of Large Scale Fields for Primordial Star Formation}\label{S:implications}

\subsection{The Power of Magnetically Mediated Outflows}

The primary effect of the magnetic field that we consider is the
generation of a magnetically mediated jet or wind.  Integrating the
vertical Poynting flux through the disk in each vertical half gives an
estimate of the magnetic luminosity available for driving each
outflow:
\beq
L_{\rm mag,1/2}\sim \bB_{\phi,s} \bB_{z,s} \Omega r^3,
\label{lmag1}
\eeq where $\bB_{\phi,s}$ and $\bB_{z,s}$ are the toroidal and
vertical fields just inside the disk surface (defined at one
scaleheight above the midplane). The field strengths we estimated in
the previous section represent fields inside the disk, so the use of
equation~(\ref{lmag1}) really requires a solution of the global
boundary value problem (e.g. R\"udiger, Elstner \& Stepinski 1995;
Reyes-Ruiz \& Stepinski 1997;1999). For present purposes, we make
crude estimates for the surface fields.  We take $\bB_{\phi,s}\sim
\bB_{\rm \phi,ave}=\alpha_{\rm ss}^{-1/2}\bB_{r,{\rm ave}}$ as
estimated in the previous section, where $\bB_{\rm \phi,ave}\sim
(3/4)\bB_{\phi}$ and $ \bB_{r,{\rm ave}} \sim (3/4)\bB_{r}$ are the
respective average values over a scaleheight, using flux freezing in
which the density falls a factor of $1/e$ between the midplane and
surface. Note that for a constant $\alpha_{\rm ss}$ disk, $v_A/c_s$ is
a constant (Blackman 1998). Then with $c_s\propto r^{-1/2}$ and
$\rho\propto r^{-3/2}$, we have $B_r\propto r^{-5/4}$, so that $L_{\rm
  mag}$ is dominated by the contribution generated at the inner disk
radius.

Now we use the above relations to estimate $\bB_{z,s}$ by appealing to
$\nabla\cdot\bbB=0$.  We integrate this over an annular wedge of the
disk extending from the midplane to the surface, with azimuthal extent
$\Delta \phi$, and with radial extent $\delta r$, where $\delta r \sim
r$ is the characteristic radial scale over which the vertical and
radial fields change significantly.  The toroidal flux does not
contribute to this integral, and by the assumption that the vertical
field vanishes at the midplane, no net flux penetrates the equatorial
plane. Then integrating $\nabla\cdot\bbB=0$ over the described volume
gives 
\beq
\Delta \phi
\left[B_{r,{\rm ave},i}h_{i}r_i-r_oB_{r,{\rm ave},i}\left(\frac{r_i}{r_o}\right)^{5/4}\left(\frac{h_{i}}{r_{i}}\right)r_o\right]
=\Delta \phi 
\frac43\left[B_{z,i}r_i^{5/4}(r_o^{3/4}-r_i^{3/4})\right], 
\eeq
where the subscripts $i$ and $o$ refer to inner and outer radii,
and we assume $h/r=h_i/r_i$, that is, the ratio is independent of $r$, which 
is a reasonable approximation (see Figs. \ref{fig:diskexam01} and \ref{fig:diskexam.3}).
In the limit $r_o \gg r_i$ and using $B_{r,{\rm ave}}=(3/4)B_r$ from above,
we obtain 
\beq
B_{z,i}/B_{r,i}= (9/16)(h/r).
\eeq
From equation~(\ref{bphi2}), $\bB_{\phi,s}=\alpha_{\rm ss}^{-1/2}B_{r,{\rm ave}}=(3/4)\alpha_{\rm ss}^{-1/2}B_{r,i}$
so that equation~(\ref{lmag1}) for the outflow power becomes
\beq
L_{\rm mag,1/2}\sim (27/64)(h/r)\alpha_{\rm ss}^{-1/2} \bB_{r}^2 \Omega r^3.
\label{lmag2}
\eeq 
The procedure for using equation~(\ref{lmag2})
to assess the value of $L_{\rm mag}$ for star-forming disks,
is to pick a particular star formation model 
(which means choosing  $\alpha_{\rm ss}$, $m_*$ and ${\dot m}_*$)
and inferring from its solution
whether the field has saturated or not (from eqs. \ref{exp} and
\ref{exp2}).  The appropriate field strength is then used in
equation~(\ref{lmag2}) to estimate the wind luminosity.

The case of a saturated field (eq.~\ref{exp2}) is the most interesting:
\begin{eqnarray}
\label{lum2a}
L_{\rm mag,1/2} & \sim & (27/64) \alpha_{\rm ss}^{1/2} \frac{Gm_* \dot{m}_*}{r} 
= 0.84 \alpha_{\rm ss}^{1/2} L_{\rm max,1/2}, \\
 & = & 92
\left(\frac{\alpha_{\rm ss}}{0.01}\right)^{1/2} \frac{m_*}{M_\odot}
\frac{{\dot m}_*}{0.01\smyr}\left(\frac{r}{10^{13}{\rm cm}}\right)^{-1}\:L_\odot,
\label{lum2}
\end{eqnarray}
where $L_{\rm max,1/2}=Gm_*{\dot m}_{*}/2r$ is the upper limit on the
magnetic luminosity through each disk plane available from accretion.
Note that $L_{\rm mag,1/2}< L_{\rm max,1/2}$ as expected, for all
reasonable values of $\alpha_{\rm ss}$ (formally $\alpha_{\rm ss} <
1$). The evolution of $L_{\rm mag}$ with $m_*$ is shown in Figure
\ref{fig:mLpwind2}a, together with the radiative luminosity from the
fiducial protostellar evolution model of TM. The wind luminosity is a
fixed fraction of the accretion luminosity from the inner disk.


Fields produced by a helical dynamo are ordered such that  
their smallest gradient scale exceeds that which would otherwise
result in their turbulent destruction before rising to the corona.
But there is an additional process in the generation
of global large scale fields in the corona that we have not discussed
above: the fields supplied by
the disk dynamo must incur further evolution
to even larger scales once in the corona. Their scale in the disk
is a few times $h$, but the needed large scale in the corona
is $\sim r$.  While this further opening or ``dynamical magnetic relaxation'' process requires further study with direct 
application to the disk-corona system, there is ample reason to believe
that it happens sufficiently fast when the supplied fields from the dynamo
are helical (as they would be from a helical disk dynamo, and further
supplemented by foot point shear):
It has long been known that helical fields relax to the largest
scale available to them in magnetically dominated environments 
(Taylor 1986) and it has been further shown recently 
that when helical magnetic fields are injected into such a magnetically
dominated region, this relaxation occurs on an Alvf\'en crossing
time (Blackman \& Field 2003). 
The speed is independent of the value of the magnetic diffusivity,
as long as the latter is not identically zero. For this reason
the dynamical relaxation in the corona would naturally allow
further opening of disk-supplied fields to global scales, 
not unlike the ``magnetic
carpet'' evolution in the solar corona (Schrijver \& Zwaan  2000).

Given the magnetic luminosity, we estimate the mass-loading and
momentum of the outflow. We assume that at large distances from the
star, the energy content of the flow is dominated by the kinetic
energy of the matter, so that $L_{\rm mag}=\dot{m}_w v_w^2/2$, where
$\dot{m}_w$ is the mass outflow rate and $v_w$ is the terminal
velocity. Theoretical models of magneto-centrifugally driven outflows
suggest that $v_w$ is about equal to the escape velocity from the
protostar, $v_{\rm esc,*}$ (e.g. Shu et al. 2000; K\"onigl \& Pudritz
2000). We write $v_w = f_v v_{\rm esc,*}$, where $f_v$ is a factor of
order unity.  Then the mass flux of the outflow is
\beq
\dot{m}_w = \frac{27}{32}\frac{\alpha_{\rm ss}^{1/2}}{f_v^2} \dot{m}_{*}\rightarrow 0.084 \dot{m}_*.
\label{mdotw}
\eeq We note that $\alpha_{\rm ss}$ is uncertain, as are the numerical
prefactors of equations (\ref{lum2}) and (\ref{mdotw}). For a massive
protostar that is approaching the Eddington limit, the mass-loading of
the outflow may be significantly enhanced.

\subsection{Feedback Effects from Outflows}

In present-day star formation, protostellar outflows contribute the
following feedback effects: (1) they prevent the accretion of gas in
the polar regions of the star-forming core that would otherwise have
joined the star and they divert a fraction of the accretion flow in
the disk (e.g. Matzner \& McKee 2000, hereafter MM); (2) they inject
mechanical energy into giant molecular clouds, which helps to keep the
clouds turbulent (Norman \& Silk 1980); (3) through the presence of
dense, outflowing gas near the star, they inhibit the reach of other
feedback effects, such as ionizing radiation (Tan \& McKee 2003b).  We
discuss the significance of these processes in the context of
primordial star formation.

\subsubsection{Star Formation Efficiency}

First, consider the expulsion of gas from the polar regions of the
core. By balancing magnetic tension and magnetic pressure gradients,
Matzner \& McKee (1999) (see also Shu et al. 1995; Ostriker 1997)
showed that far from the star the angular distribution of the momentum
in a radial hydromagnetic wind can be approximated by \beq
\frac{dp_w}{d\Omega} = \frac{p_w}{4\pi} \frac{1}{{\rm
    ln}(2/\theta_0)(1+\theta_0^2 - \mu^2)},
\label{pangular}
\eeq where $\mu={\rm cos}\theta$ and $\theta_0$ is a small angle,
which is estimated to be $\sim 0.01$ for winds from low-mass stars
(Matzner \& McKee 1999), but may be somewhat larger for winds from
massive stars.  We adopt this value in the primordial case (our
results are not particularly sensitive to this choice).
Equation~(\ref{pangular}) is also a reasonable description for the
angular distribution of the mass flux in the wind, assuming the
terminal velocity (far from the star, but before deceleration by the
ambient gas) is independent of $\theta$.  

To evaluate the efficiency of star formation from a core (i.e. the
fraction of mass that forms a star) $\ecore$, we follow MM and find
the angle, $\theta_{\rm esc}\equiv{\rm cos}^{-1}\mu_{\rm esc}$, where
the wind sweeps up core material to the surface escape speed of the
core, $v_{\rm esc,c}$.  The analysis assumes thin, radiative shocks,
purely radial motion and monopole gravity. We consider cores with
angular mass distributions of the form $dM/d\Omega =(1/4\pi) Q(\mu) M$
with $Q(\mu)=(1-\mu^2)^n / \int_0^1 (1-\mu^2)^n \:d\mu$, and evaluate
models with $n$, ranging from $0$, isotropic, to $4$, which describes
a flattened distribution that mimics the effect of some rotational
support. In the notation of MM, the escape condition is given by
$(1+\theta_0^2 - \mu_{\rm esc}^2) Q(\mu_{\rm esc}) = m_*/(X M)$, where
$X=0.132 c_g [{\rm ln}(2/\theta_0) / {\rm ln} 200] v_{\rm esc,c,5} [
(p_w/m_*)/ 40{\rm km\:s^{-1}} ]^{-1}$, and $c_g$ is a factor of order
unity that accounts for the effects of gravity on shock propagation.
Using the results of MM, we estimate $c_g\simeq4.6$, for steady winds
that decouple from the swept-up shell at the core edge in a core with
$k_\rho = 20/9$ (which is the fiducial value of TM). These cores have
$v_{\rm esc,c}=3.22 (M/1000M_\odot)^{-1/7} K'^{5/7}\:{\rm km/s}$.  We
ignore the influence of material beyond the core ``edge'', which is
partly balanced by our neglect of the wind's influence on the shell
beyond this point.  The efficiency is given by \beq \ecore =
\frac{1}{1+(\dot{m}_w/\dot{m}_*)(1-\phi_w)} \int_0^{\mu_{\rm esc}} Q
\:{\rm d}\mu,
\label{ecore}
\eeq where $\phi_w\equiv \int_0^{\mu_{\rm esc}} P\:{\rm d}\mu$ and
$P(\mu)$ is the dimensionless force distribution in angle. Note that
unlike $\esd$, which was used in equation~(\ref{eq:mds}), this definition
of efficiency includes the effect of diversion of material from the
disk into the outflow. 
However, since $f_w\equiv\dot{m}_w / \dot{m}_*\ll1$, $\ecore\simeq \esd$.

In summary, we are assuming that the properties of the protostellar
outflow, and how they evolve with $m_*$, do not depend on $\ecore$. In
reality, as $\ecore$ and $\dot{m}_*$ become smaller, then the
protostellar evolution changes, e.g. the stellar growth time becomes
longer relative to the Kelvin-Helmholz time, so that the star would
tend to be smaller. However, since $p_w$ is integrated over the
stellar evolution, it should not be so sensitive to these changes.

\begin{figure}[h]
\plotone{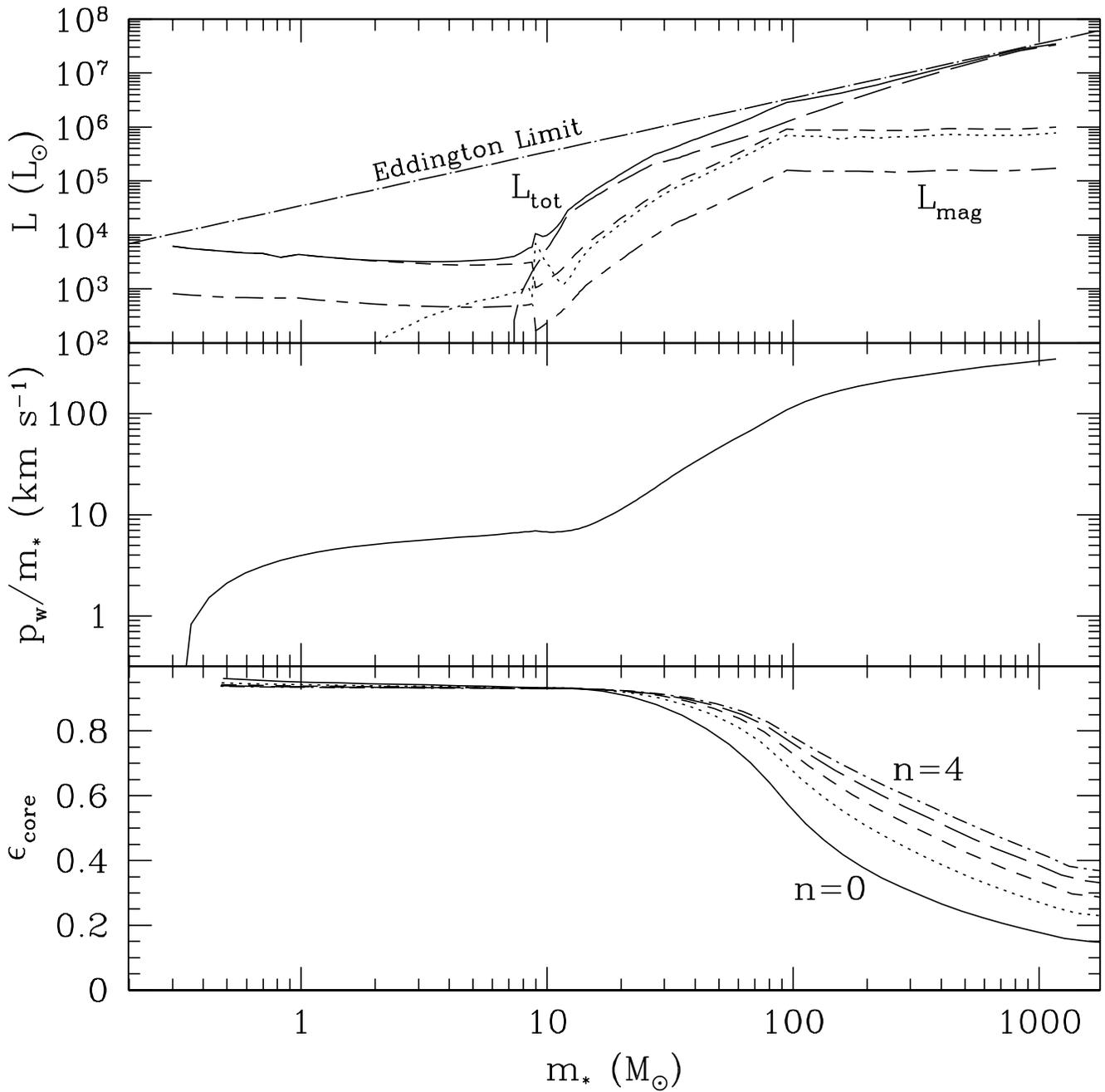}
\caption{
\label{fig:mLpwind2}
\footnotesize
Evolution of protostellar luminosities, cumulative specific outflow
momentum, and the resulting star formation efficiency for the fiducial
$f_{\rm Kep}=0.5$ and $K'=1$ case, with $\alpha_{\rm ss}=0.01$. {\it
  (a) Top panel:} Evolution of wind luminosity, $L_{\rm mag}$, is
shown by the dashed-long-dashed line. Also shown are the total
bolometric luminosity from the protostar (solid line) including
contributions from the inner ($r<10r_*$) accretion disk (dotted line),
stellar accretion luminosity including from direct spherical accretion
and boundary layer accretion (dashed line), and internal protostellar
luminosity (long-dashed line) (see TM for more details). The Eddington
luminosity is shown by the dot-long-dashed line.
{\it (b) Middle panel:} Cumulative
momentum of the protostellar outflow divided by the stellar mass. This
estimate assumes that the wind speed is the escape velocity from the
protostellar surface and that at large distances from the star the
energy content of the flow is dominated by kinetic energy. 
{\it (c) Bottom panel:}
Efficiency of star formation due to erosion of the gas core by the
outflow. The density profile of the initial core is specified by
$dM/d\Omega =(1/4\pi) Q(\mu) M$, with $\mu = {\rm cos} \theta$ and
$Q(\mu)=(1-\mu^2)^n / \int_0^1 (1-\mu^2)^n \:d\mu$. Solid line is
$n=0$ (isotropic core), dotted is $n=1$, dashed is $n=2$, long-dashed
is $n=3$, dot-dashed is $n=4$.}
\end{figure}

We show the evolution of $p_w/m_*$ with $m_*$ for our fiducial model
in Figure \ref{fig:mLpwind2}b. Initially $r_*$ is approximately
constant and the specific momentum of the wind grows only because of
the increasing mass and escape velocity from the star. Then above
$m_*\simeq 20M_\odot$, the star starts shrinking towards the main
sequence and there is a jump in the momentum input. This marks the
start of the transition to lower efficiency star formation from the
core (Fig. \ref{fig:mLpwind2}c).  This transition may have important implications for the final
mass of the star, and will be examined in more detail in a future
paper.  Note that this efficiency is not particularly sensitive to the
parameter $\theta_0$ when the critical angle for ejection of gas from
the core is $\gtrsim \theta_0$. That is, mass limits to star formation
resulting from this process do not depend on $\theta_0$ so long as
$\theta_0\ll 1$.

\subsubsection{Feedback on the Larger-Scale Cloud and Break-Out from the Dark Matter Halo}

We have seen that the protostellar outflow is eventually able to break
free of the core. With the distribution of momentum given by equation
(\ref{pangular}), this is particularly true near the rotation axis,
since about half of the initial outflow mass is concentrated inside
$\theta_0$. A $100\sm$ star releases a wind with total momentum of
about $10^4 M_\odot\:{\rm km/s}$, with about a quarter of this inside
$\theta_0$ of each jet, for which the injected mass is about $2\sm$.
Each central jet region can then sweep up $\sim 500\sm$ to about
$5\:{\rm km/s}$, which is the approximate sound speed and escape speed
of the gas in the vicinity of the dark matter mini-halo. The fiducial
core of TM has a total virialized baryonic mass of $\sim 10^3\sm$
(based on the simulation of Abel et al. 2002), with a few $\times
10^4\sm$ of gas still infalling within a couple of hundred parsecs.
For reasonable values of $\theta_0\lesssim 0.1$, and even with
allowance for turbulent entrainment of matter at the sides of the jets
and lateral spreading, the outflows should easily be able to eject
some gas from the dark matter halo. Thus we expect the extent of
mechanical feedback to reach beyond the immediate environment of the
star-forming core.

The physical implications of this feedback are difficult to assess,
since radiative feedback from the massive protostar is also acting on
the cloud during this time. If outflows do escape from the halo, then
they will start to pollute the intergalactic medium with their
magnetic fields. This is only one possible mechanism for producing
such fields. Others include field generation in shocks caused by
supernovae or ionization fronts (Subramanian et al. 1994; Gnedin et
al. 2000), and these mechanisms should have a longer reach than the
protostellar outflows. Outflows from AGN are also likely to be
important, although these may be more sparsely distributed than
Population~III stars.

\subsubsection{Confinement of Radiative Feedback by Protostellar Outflows}

The presence of dense, outflowing gas near the protostar can suppress
the influence of radiative feedback. For example, the propagation of
ionizing photons should be impeded. The characteristic density at the
base of the flow region near the star is 
\beq n_{\rm H,base} = 4.0
\times 10^{11} f_v^{-1} \left(\frac{f_r}{2}\right)^{-2}
\left(\frac{f_w}{0.1}\right)
\left(\frac{\dot{m}_*}{10^{-3}\smyr}\right) \left(\frac{r_*}{10\:{\rm
      R_\odot}}\right)^{-3/2}
\left(\frac{m_*}{100\:M_\odot}\right)^{-1/2}\:{\rm cm^{-3}},
\label{nbase}
\eeq where $f_w\equiv \dot{m}_w/\mds$ 
and we have assumed the wind is launched from the inner part of
the disk from an annulus of inner radius $f_r r_* = 2r_*$,
and outer radius $2 f_r
r_*$. 

Assuming an isotropic flux, the ionizing luminosity necessary to
balance recombinations and fully ionize the gas, is given by
\beq 
\frac{S_{\rm crit}}{4 \pi} = \int_{f_r r_*}^{2f_r r_*} \alpha^{(2)} n_{\rm H,base} n_e r^2 dr,
\label{flux}
\eeq
where $\alpha^{(2)}=2.62\times 10^{-13}\:{\rm s^{-1}}$ is the radiative
recombination rate to the excited states of hydrogen at $10^4\:{\rm
  K}$ and $n_e$ is the electron number density. We assume singly ionized He, so that
\beq
S_{\rm crit}=2.9 \times 10^{47} f_v^{-2} \left(\frac{f_r}{2}\right)^{-1} \left(\frac{f_w}{0.1}\right)^2 \left(\frac{\dot{m}_*}{10^{-3}\smyr}\right)^2 \left(\frac{m_*}{100\:M_\odot}\right)^{-1}\:{\rm ph\:s^{-1}}.
\label{Scrit}
\eeq Models of Population~III stellar structure (Tumlinson \& Shull
2000; Baraffe, Heger, \& Woosley 2001; Bromm, Kudritzski, \& Loeb
2001; Schaerer 2002) yield zero age main sequence (ZAMS) ionizing
luminosities that can be approximated via $S=9.42\times
10^{49}(m_*/30\sm)^{2.5}\:{\rm ph\:s^{-1}}$ for ($12<m_*/\sm<40$),
$S=9.17\times 10^{49}(m_*/100\sm)^{1.7}\:{\rm ph\:s^{-1}}$ for
($40<m_*/\sm<120$) and $S=3.75\times 10^{50}(m_*/300\sm)^{1.2}\:{\rm
 ph\: s^{-1}}$ for ($120<m_*/\sm<1000$). The critical ZAMS mass that can
ionize the outflow from a protostar accreting at the fiducial rate
(Eq.~\ref{eq:mds}) is then given by 
\beq m_{\rm *,crit}=34 f_v^{-0.46}
\left(\frac{f_r}{2}\right)^{-0.23} \left(\frac{f_w}{0.1}\right)^{0.46}
\esd^{0.66} K'^{0.98} \sm,
\label{mcrit}
\eeq so long as $12<m_{\rm *,crit}/\sm<40$. Similar relations can be
derived for when $m_{\rm *,crit}$ is in a higher mass range.

Primordial protostars only reach the ZAMS at masses around $100\sm$
(Omukai \& Palla 2001; 2003; TM). Before this the ionizing luminosity
is initially very small because the protostar is large and relatively
cool.  In the fiducial case the ionizing luminosity, including the
contribution from accretion luminosity, then increases rapidly to
become equal to the ZAMS value at about $40\sm$ (Figure 7 of TM).
Therefore, we expect the outflow to be able to confine ionizing
radiation for much of the pre-main sequence phase of protostellar
evolution up to about $40\sm$, but not after that. These results have
implications for feedback models that are based on the ionization of
the protostellar disk (e.g. Hollenbach et al. 1994). In present-day
massive star formation this process of confinement by the outflow can
create extremely compact \ion{H}{2} regions around massive protostars
(Tan \& McKee 2003b). A more complete analysis of the feedback effects
in the primordial case, including those from FUV photons, is beyond
the scope of this paper and is deferred to a future study.

\section{Conclusions}\label{S:conclusions}

We have presented calculations of the radial structure of primordial
protostellar disks, using realistic models for the rate of accretion
and the size of the protostar (Tan \& McKee 2003a).  We have followed
the evolution of disk properties as the protostar grows in mass and
changes in size. These disks differ from those around present-day
protostars, because of the lack of cooling due to dust and because of
the high rates at which they are fed matter from the collapsing core.
The high accretion rates and large protostellar sizes require
inclusion of the ionization energy in the structure equations, the
effect of which is to reduce the disk temperatures and luminosities,
particularly for disks around protostars with masses $\lesssim 20\sm$.
We argue that fragmentation in the disks is unlikely to occur, if
local gravitational instabilities can provide an effective viscosity
of the magnitude seen in the simulations of Gammie (2001).

We have shown that in situ dynamos can produce dynamically significant
large scale magnetic fields in primordial star-forming disks. For
present-day disks, such fields would likely be important for driving
outflows in all star-forming systems.  For the primordial case, the
disks of the lowest mass protostars would not have enough time to
generate strong fields, but the majority of the star-forming parameter
regime at higher masses leads to strong fields even for very
conservative assumptions about the initial seed magnetic field.  Thus
models of primordial star formation should include the dynamical
influence of magnetic fields.

We emphasize that for these conclusions to be valid, gravitational
instability, rather than the MRI, must provide the initial source of
turbulence in the disk.  Even though the ionization fraction is high,
the seed field is below the critical value required at the mode of
fastest growth. This contrasts with the situation in present-day
protostellar disks.  A second requirement is that the turbulence in
the disk must generate helicity.  It is important for this stage of
the evolution to be investigated with numerical simulations that can
resolve the helicity.

If large scale magnetic fields can be amplified to their saturation
strength, then their main effect is to drive a collimated wind. This
reduces the star formation efficiency by diverting material from the
accretion disk and by sweeping-up ambient gas from the polar regions,
that otherwise would have collapsed. The star formation efficiency is
significantly reduced once the protostar reaches about $100\sm$.  The
outflows are probably strong enough to eject matter from the dark
matter mini-halo and to start magnetizing the intergalactic medium.
The presence of outflows also increases the density distribution of
gas near the star in the region above the accretion disk. The full
effects of this have not yet been worked out, but it appears likely
that the propagation of ionizing photons is inhibited, at
least while the protostar is in the earlier stages of contraction
towards the main sequence. These results will be applied to a
more detailed study of protostellar feedback in a forthcoming paper.


\acknowledgements JCT is supported by a Spitzer-Cotsen fellowship from
Princeton University and NASA grant NAG5-10811.  EGB acknowledges
support from DOE grant DE-FG02-00ER54600.  We thank C. McKee, V.
Pariev, and K. Omukai for discussions.

\end{document}